\documentclass[10pt,aps,prd,showpacs,superscriptaddress,twocolumn]{revtex4}
\usepackage{amsfonts}
\usepackage{amssymb}
\usepackage{amsmath}
\usepackage{MnSymbol}
\usepackage{graphicx}
\usepackage{epstopdf}

\date{\today}

\begin{document}

\title{Quantum versus classical instability of scalar fields in 
curved backgrounds}

\author{Raissa F.\ P.\ Mendes}
\email{rfpm@ift.unesp.br}
\affiliation{Instituto de F\'\i sica Te\'orica, Universidade 
Estadual Paulista, Rua Dr. Bento Teobaldo Ferraz 271, 01140-070, 
S\~ao Paulo, S\~ao Paulo, Brazil}

\author{George E.\ A.\ Matsas}
\email{matsas@ift.unesp.br}
\affiliation{Instituto de F\'\i sica Te\'orica, Universidade 
Estadual Paulista, Rua Dr. Bento Teobaldo Ferraz 271, 01140-070, 
S\~ao Paulo, S\~ao Paulo, Brazil}

\author{Daniel A.\ T.\ Vanzella}
\email{vanzella@ifsc.usp.br}
\affiliation{Instituto de F\'\i sica de S\~ao Carlos,
Universidade de S\~ao Paulo, Caixa Postal 369, 13560-970, 
S\~ao Carlos, S\~ao Paulo, Brazil}

\begin{abstract}
General-relativistic stable spacetimes can be made unstable under 
the presence of certain nonminimally coupled free scalar fields. 
In this paper, we analyze the evolution of linear scalar-field
perturbations in spherically symmetric spacetimes and compare the
classical stability analysis with a recently discussed quantum field
one. In particular, it is shown that vacuum fluctuations lead to 
natural seeds for the unstable phase, whereas in the classical
framework the presence of such seeds in the initial conditions
must be assumed. 
\end{abstract}


\maketitle

%%%%%%%%%%%%%%%%%%%%%%%%%%%%%%%%%%%%%%%%%%%%%%%%%%%%%%%%%%%%%%%%%%
\noindent
\textbf{Introduction}:
It was shown that certain well-behaved 
spacetimes can induce an exponential growth of the vacuum 
energy density of some nonminimally coupled free scalar fields~\cite{lv}. 
Particular astrophysically inspired realizations of this 
mechanism were explored in Refs.~\cite{lmv,lmmv}. As the 
instability sets in, the system is driven to a new equilibrium 
state, generically inducing a burst of free scalar 
particles~\cite{llmv}. This quantum field effect has a 
classical counterpart, as remarked in Ref.~\cite{panietal}. 
There, the authors discuss the end state of the (classical) 
instability, and provide evidence that, for a certain 
range of field-to-curvature couplings, the system 
evolves to a ``scalarized'' final configuration 
(see also Refs.~\cite{harada97,novak}).

In this paper we provide a more rigorous formulation 
of the relationship between the quantum and classical descriptions 
of the instability. For this purpose, we analyze the evolution 
of classical perturbations in a regular spherically symmetric spacetime
within a quasinormal mode formalism based on the Laplace transform 
approach. This approach, which is often used to analyze the 
evolution of stable perturbations~\cite{leaver86,nollert92,reviewQNM} 
is adapted here to 
unstable ones. We then show the similarities and differences
between the quantum and classical descriptions 
and how quantum fluctuations can be simulated by classical 
perturbations of a given ``small" amplitude. 
We set $c = G = 1$. 
\\

%%%%%%%%%%%%%%%%%%%%%%%%%%%%%%%%%%%%%%%%%%%%%%%%%%%%%%%%%%%%%%%%%%

\noindent
\textbf{Quantum approach to instability - a brief review}:
Let us assume a real massless free scalar field $\Phi$,
on a spacetime $(\mathbb{R}^4, g_{ab})$, governed by the 
field equation 
\begin{equation} \label{KG}
	(-\nabla^a \nabla_a  + \xi R) \Phi = 0,
\end{equation}
where $R$ is the scalar curvature and $\xi \in \mathbb{R}$.
Throughout this paper we restrict our attention to asymptotically flat, 
spherically symmetric spacetimes possessing no event horizons or
singularities. Moreover, let the spacetime be Minkowski-like in the
past, $(\mathbb{R}^4,\eta_{ab})$, and static in the future, 
$(\mathbb{R}^4,g_{ab})$, with $g_{ab}$ such that
\begin{equation} \label{metric}
	ds^2 = - e^{2\Xi (r)} dt^2 + e^{2\Lambda (r)} dr^2 + r^2(d\theta^2 
	+ \sin^2\theta d\phi^2 ),
\end{equation}
where $\Xi(r),\Lambda(r) \overset{r\sim 0}{\sim} r^2 + {\cal O} (r^4)$  
are bounded continuous functions with  $\Xi(r),\Lambda(r) \overset{r\to\infty}{\to} 0$. 

References ~\cite{lv,lmv,lmmv,llmv} analyzed the case of a quantum
field $\hat{\Phi}$, which satisfies Eq.~(\ref{KG}), being stable on 
$(\mathbb{R}^4,\eta_{ab})$ but unstable on $(\mathbb{R}^4,g_{ab})$. 
In this setting, if the field is in the no-particle state 
$|0_\textrm{in}\rangle$ as described by static observers in 
$(\mathbb{R}^4,\eta_{ab})$, then its vacuum fluctuations 
%(and, consequently, the expectation value of its energy-momentum tensor)
suffer an exponential amplification in time during the unstable phase:
\begin{equation} \label{evphi2}
	\langle 0_\textrm{in}| \hat{\Phi}^2 |0_\textrm{in}\rangle 
	\sim
	\frac{\hbar \kappa e^{2\bar{\Omega}t}}{8 \pi\bar{\Omega}} 
	\left( \frac{\psi_{\bar{\Omega}0}(r)}{r} \right)^2 [1+ {\cal O}
	(e^{-\epsilon t})].
\end{equation}
Here, $\psi_{\Omega l}(r)$, with $\Omega > 0$ and $l \in \mathbb{N}$, 
obeys
\begin{equation} \label{psieq2}
	[- d^2/dx^2 + V^{(l)}_{\textrm{eff}}(r)] \psi_{\Omega l} [r(x)]
	= -\Omega^2 \psi_{\Omega l} [r(x)],
\end{equation}
where the effective potential is given by
\begin{equation} \label{pot}
	V^{(l)}_{\textrm{eff}}(r) = e^{2\Xi} \left(\xi R + \frac{l(l+1)}{r^2}
	\right) + \frac{e^{2(\Xi-\Lambda)}}{r} \left( \frac{d\Xi}{dr} 
	-\frac{d\Lambda}{dr} \right)
\end{equation}
and $x \in [0,+\infty)$ is defined as
\begin{equation}
x(r) \equiv\int_0^r  e^{\Lambda(r') - \Xi (r')} dr'.
\label{x}
\end{equation}
Moreover, proper behavior of the field at the origin and infinity
demands
\begin{equation}
\psi_{\Omega l} [r(x)]_{x = 0} =0, \qquad
\psi_{\Omega l} [r(x)]|_{x\to +\infty} \sim
e^{-  \Omega \, x}, 
\label{boundarycond}
\end{equation}
while normalization requires
\begin{equation} \label{normpsi2}
	\int_0^{+\infty}{dx \psi^*_{\Omega l} \psi_{\Omega' l}} = 
	\delta_{\Omega \Omega'}.
\end{equation}
In Eq.~(\ref{evphi2}),
$\epsilon$ is some positive constant, $-\bar{\Omega}^2$ 
is the lowest negative eigenvalue of  
$- d^2/dx^2 + V^{(0)}_{\textrm{eff}}(x)$, and $\kappa = {\rm const} \sim 1$ 
%is a multiplying factor of order unity carrying information about the
depends on the transition details to the unstable phase.

For a minimally coupled field ($\xi=0$), it is possible to show 
that the operator $- d^2/dx^2 + V^{(l)}_{\textrm{eff}}(x)$ has a 
purely positive spectrum and thus no solutions of 
Eq.~(\ref{psieq2}) satisfying Eq.~(\ref{boundarycond}) exist 
(see the Appendix). However, for 
nonminimally coupled fields, the effective potential can be made
sufficiently negative to allow the same operator to possess an 
additional negative (discrete) spectrum. This is the hallmark
of the instability. See Refs.~\cite{lv,lmv,lmmv,llmv} for a complete 
discussion on the ``vacuum awakening effect'' (and Ref.~\cite{lima} 
for a rigorous discussion on the quantization 
of unstable linear fields in globally static spacetimes).
\\

%%%%%%%%%%%%%%%%%%%%%%%%%%%%%%%%%%%%%%%%%%%%%%%%%%%%%%%%%%%%%%%%%%
\noindent
\textbf{Connection with classical approach to instability}:
We now investigate the classical counterpart of the 
quantum instability described above.
Consider the action $S_\Psi$
describing some matter field $\Psi$ defined on a spacetime 
ruled by the Einstein-Hilbert action $S_{\rm EH}$.
Variation of $S_{\rm EH} + S_\Psi$ with respect to the 
metric gives
\begin{equation} \label{EE0}
	G_{ab} = 8 \pi T^\Psi_{ab},
\end{equation}
where 
$ T_{ab}^\Psi = -({2}/{\sqrt{-g}}) {\delta S_\Psi}/{\delta g^{ab}}$
and $G_{ab}$ is the Einstein tensor.
The line element~(\ref{metric})
is assumed to be a solution of Eq.~(\ref{EE0}) for some matter
distribution.

Next, let us perturb the system by introducing a free 
scalar field $\Phi$ ruled by Eq.~(\ref{KG}). Then, Eq.~(\ref{EE0}) becomes
\begin{equation} \label{EE}
	G_{ab} = 8 \pi ( T^\Psi_{ab} + T^\Phi_{ab}),
\end{equation}
where
\begin{align}
	T^\Phi_{ab} &= (1-2\xi) \nabla_a \Phi \nabla_b \Phi + 
	\xi \Phi^2 R_{ab}- 2 \xi \Phi \nabla_a \nabla_b \Phi
	\nonumber \\
	&+ (2\xi-1/2)  
	[ \nabla_c \Phi \nabla^c \Phi	+ \xi R \Phi^2] 
	\,g_{ab}.
	\label{Tmunu}
\end{align} 

Inspired by the previous section where the quantum field was chosen to
be in a suitable vacuum state, we aim to solve the classical field 
equations up to linear-order perturbation over the null-scalar-field 
configuration. Then, let us define
$g_{ab} \equiv g^{(0)}_{ab} + g^{(1)}_{ab}, \;\;\; 
\Phi  \equiv \Phi_{(0)} + \Phi_{(1)},$
where  $\Phi_{(0)}=0$ and $\Phi_{(1)}$ is small in the sense that it 
engenders a small perturbation $g^{(1)}_{ab}$ with respect to the 
unperturbed background metric $g^{(0)}_{ab}$ given by Eq.~(\ref{metric}). 
Because $ T_{ab}^\Phi$ has a quadratic dependence in $\Phi$, we 
conclude that at first-order perturbation $g_{ab} = g^{(0)}_{ab}$ 
is a solution of Eq.~(\ref{EE}), while $\Phi_{(1)}$ evolves according to
\begin{equation} \label{KG2}
	(-\nabla^a \nabla_a + \xi R) \Phi_{(1)} = 0
\end{equation}
on the fixed background $( \mathbb{R}^4, g^{(0)}_{ab} )$.

In contrast to the quantum case where vacuum fluctuations 
automatically trigger the
exponential growth of  $\langle 0_\textrm{in} | \hat{\Phi}^2 | 0_\textrm{in} \rangle$
while keeping 
$\langle 0_\textrm{in} | \hat \Phi | 0_\textrm{in} \rangle =0$, in the classical 
context we shall postulate that at some instant, say $t=0$,
some external agent drives $\Phi (t,{\bf x})$  
out of its initial equilibrium state such that 
$\Phi (t,{\bf x}) \neq 0$ itself for $t>0$. Due to the 
spherical symmetry of the background spacetime~(\ref{metric}), 
we decompose $\Phi_{(1)}$ as
\begin{equation} \label{dphi}
	\Phi_{(1)} (t,r,\theta,\phi) = \sum_{l=0}^{\infty} 
	\sum_{\mu=-l}^l \frac{\chi_{l\mu}(t,r)}{r} Y_{l \mu}(\theta,\phi),
\end{equation}
where the initial conditions are defined by specifying
$\chi_{l\mu}(t,r)$ and  $\partial_t\chi_{l\mu}(t,r)$ at $t=0$. 
In order to handle the initial conditions and establish a 
clear connection between the quantum analysis and the 
one using quasinormal modes, it will prove 
convenient to adapt the Laplace transform 
approach~\cite{nollert92} to our case.

Let us define the Laplace transform of $\chi_{l\mu}(t,r)$ with 
respect to the time coordinate as
\begin{equation} \label{LT}
	\tilde{\chi}_{l\mu}(s,r) \equiv \int_0^{+\infty}{e^{-st} \chi_{l\mu}(t,r) dt}, 
	\;\;\;s \in \mathbb{C},
\end{equation}
in some domain $\Re(s) > \gamma$ where $\tilde{\chi}_{l\mu}(s,r)$ is analytic. 
Here $\gamma$ is chosen so that  $|\chi_{l\mu}(t,r)| \leq M e^{\gamma t}$ 
whenever $t>t_0$, for some $t_0, M \in \mathbb{R}^+ $. 
From Eqs.~(\ref{KG2})-(\ref{LT}), it follows 
that $\tilde{\chi}_{l\mu}(s,r)$ obeys
\begin{equation} \label{chieq}
	-\partial_x^2\tilde{\chi}_{l\mu}(s,r) + (s^2 + V^{(l)}_{\textrm{eff}}) \,
	\tilde{\chi}_{l\mu}(s,r) =  \mathcal{I}_{\,l\mu}(s,r),
\end{equation}
where
\begin{equation} \label{I}
	\mathcal{I}_{\, l\mu}(s,r) \equiv [s \chi_{l\mu} (t,r) + \partial_t\chi_{l\mu} (t,r)]_{t = 0}
\end{equation}
is fixed by the initial conditions. Inspired by the quantum 
case where the instability is triggered by vacuum fluctuations which drop fast at 
infinity [see Eq.~(\ref{evphi2}) with $\psi_{\bar \Omega 0}$ 
obeying Eq.~(\ref{boundarycond})], we consider here 
that the system is perturbed by a classical seed 
localized in space. Thus, we assume that $\chi_{l\mu}(t,r)$ and 
$\partial_t \chi_{l\mu}(t,r)$ have compact support 
as functions of $r$ at $t=0$ in which case 
$\mathcal{I}_{\, l\mu}(s,r)=0$ for $r>\ell = {\rm const}$. 

As a consequence of our localized initial condition assumption, 
we have for  large enough $r$ that
\begin{equation}
	|\tilde{\chi}_{l\mu}(s,r)| 
	\leq \int_{x(r)-x(\ell )}^{+\infty} M e^{-( \Re (s)- \gamma)t} dt, 
	\label{aux}
\end{equation}
where we have used Eq.~(\ref{LT}) and the causal propagation property
of Eq.~(\ref{KG2}). Hence, after performing the  
integration in Eq.~(\ref{aux}), we conclude that
\begin{equation}
\lim_{r\to +\infty} \tilde{\chi}_{l\mu}(s,r) =0.
\label{chi(s)infinito}
\end{equation}
The detailed form of $\tilde{\chi}_{l\mu}(s,r)$ will depend on
Eq.~(\ref{chieq}). A general solution of Eq.~(\ref{chieq}) can be 
cast as
\begin{equation} \label{chiG}
	\tilde{\chi}_{l\mu}(s,r) = \int_0^{+\infty} {G_l(s;r,r') 
	\mathcal{I}_{\, l\mu}(s,r') dx'},
\end{equation}
where $r'\equiv r(x')$ and $G_l(s;r,r')$ satisfies
\begin{equation} \label{Green}
	-\partial_x^2 G_l(s;r,r') + (s^2 + V^{(l)}_{\textrm{eff}})
	G_l(s;r,r') = \delta (x-x').
\end{equation}
Any solution of Eq.~(\ref{Green}) can be written as
\begin{equation} \label{G_def}
	G_l(s;r,r') = f_l^-(s,r_<) f_l^+(s,r_>) / W_l(s),
\end{equation}
where $f_l^\pm (s,r)$ are two linearly independent solutions
of the homogeneous equation 
\begin{equation} \label{feq}
	-\partial_x^2 f_l^\pm (s,r) + (s^2 + V^{(l)}_{\textrm{eff}})
	f_l^\pm (s,r) = 0
\end{equation}
with $r_< \equiv \min (r,r')$ and $r_> \equiv \max (r,r')$.
Here, 
$$
W_l(s)\equiv 
f_l^+ (s,r) \partial_x f_l^-(s,r) - f_l^-(s,r) \partial_x f_l^+(s,r).
$$ 
We note that 
$G_l(s;r,r')$ is not affected by rescaling $f^\pm_l (s,r)$ through any 
(nonzero) multiplicative constant.

The Green function $G_l(s;r,r')$ is completely specified by 
Eq.~(\ref{G_def}) by imposing  proper boundary conditions
to $f_l^\pm(s,r)$. Equation~(\ref{chi(s)infinito}) combined 
with Eqs.~(\ref{chiG}) and~(\ref{G_def}) leads to
\begin{equation} \label{bc1}
	f_l^+(s,r)|_{r \to +\infty} \sim e^{-sx}
\end{equation}
for $\gamma \geq 0$.
In addition, the regularity condition 
imposed to $\Phi_{(1)}$ at the origin 
demands $\chi_{l\mu}(t,r)|_{r = 0}=0$ and, thus,
\begin{equation} \label{bc2}
	f_l^-(s,r)|_{r = 0} = 0.
\end{equation}
Note that for large enough $|s|$,
\begin{equation} \label{Gapprox}
	G_l(s;r,r') \sim e^{-sx_>} \sinh(sx_<)/s.
\end{equation}
 
Eventually, $\chi_{l\mu}(t,r)$ is recovered through the inverse 
Laplace transform %\cite{smith}
\begin{equation} \label{ILT}
	\chi_{l\mu}(t,r) = \frac{1}{2\pi i}
	\int_{\kappa - i\infty}^{\kappa + i\infty}{e^{st}
	\tilde{\chi}_{l\mu}(s,r) ds},
\end{equation}
where $\kappa > \gamma$, and $\tilde{\chi}_{l\mu}(s,r)$ is given in
Eq.~(\ref{chiG}). Then, $\Phi_{(1)}$ is 
straightforwardly obtained from Eq.~(\ref{dphi}).

\begin{figure}[t]
\includegraphics[width=8cm]{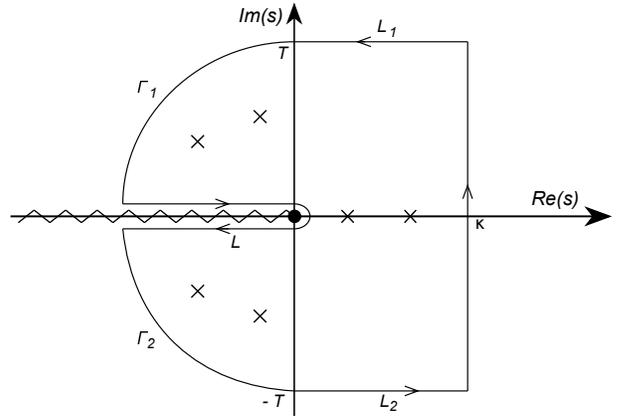}
\caption{The singularity 
structure of $\tilde{\chi}_{l\mu}(s,r)$ in the
plane $s \in \mathbb{C}$ and the integration 
contour chosen to calculate ${\chi}_{l\mu}(t,r)$ ($t>0$)
are exhibited. The logarithmic singularity is at 
the origin and the corresponding branch cut is set on 
the negative real axis. The poles of $\tilde{\chi}_{l\mu}(s,r)$ 
are represented by $\times$ symbols. We focus on
those with $s_0 > 0$ which are associated to instability.}
\label{fig:contour}
\end{figure}

For $t \geq 0$ it is convenient to extend  
$\tilde{\chi}_{l\mu}(s,r)$ to $\Re (s) \leq \gamma$ in order to use 
the residue theorem to calculate Eq.~(\ref{ILT}). The extension 
of $\tilde{\chi}_{l\mu}(s,r)$ to the remaining complex plane raises
poles including the ones which codify the instabilities in which we are 
interested. The singularities of $\tilde{\chi}_{l\mu}(s,r)$ 
in the region $\Re (s) \leq \gamma$ come from the Green 
function $G_l(s;r,r')$, since $\mathcal{I}_{\, l\mu}(s,r)$ is an entire function 
of $s \in \mathbb{C} $ [see Eq.~(\ref{chiG})]. These, in turn, can 
be traced back  either to singularities of $f_l^\pm (s,r)$ or to 
zeros of the Wronskian [see Eq.~(\ref{G_def})]. The former will 
depend on global properties of the effective potential~\cite{newton}. 
For $V^{(l)}_{\rm eff}$ associated with compact objects,  
$f_l^+(s,r)$ will possess a logarithmic singularity at 
$s=0$~\cite{chingD95}. Moreover, the zeros of the Wronskian, $W_l(s_0)=0$,  
will give rise to simple poles of $\tilde{\chi}_{l\mu}(s,r)$ at $s=s_0$ 
provided that $dW_l(s)/ds|_{s=s_0}\neq 0$ [see below Eq.~(\ref{fnormalization})]. 
Then, $f_l^-(s_0,r)$ and $f_l^+(s_0,r)$ are linearly dependent 
functions and  can be assumed to be equal with no loss of generality: 
$f_l^\pm(s_0,r) \equiv f_l(s_0,r)$. Figure~\ref{fig:contour} 
illustrates the singularity pattern of $\tilde{\chi}_{l\mu}(s,r)$.

We calculate Eq.~(\ref{ILT}) for $t>0$ through the residue 
theorem with integration contour shown in Fig.~\ref{fig:contour}:
\begin{eqnarray}
\int_{\kappa-i\infty}^{\kappa+i\infty} {e^{st} \tilde{\chi}_{l\mu} (s,r) ds}
\!\!\!\!\!\!
&&= 
2\pi i \sum_{\rm poles} {\textrm{Res}} \; [e^{st} \tilde{\chi}_{l\mu} (s,r)]  
\nonumber \\
&&-
\int_{L_1 \Gamma_1 L \Gamma_2 L_2} 	\!\!\! {e^{st} \tilde{\chi}_{l\mu} (s,r) ds}.
\label{int_scheme}
\end{eqnarray}
The $\Gamma_1, \Gamma_2$ and $L$ contributions are well studied in the 
literature~\cite{comment} (see also, e.g., Refs.~\cite{leaver86,andersson}).  
The $L_1$, $L_2$ contributions which appear as a consequence 
of the existence of poles with $\Re (s_0)>0$ can be seen to vanish 
by inserting Eqs.~(\ref{I}) and~(\ref{Gapprox}) in Eq.~(\ref{chiG})
and noting that we end up with a sum of two integrals corresponding 
to both terms of Eq.~(\ref{I}). For $T \equiv |\Im (s)| \to +\infty$, 
one of them goes to zero as $1/T$ while the other one vanishes 
as a result of the rapid oscillation of the integrand in this limit. 

Then, by using Eq.~(\ref{int_scheme}) in Eq.~(\ref{ILT}) we have 
\begin{equation}
	\chi_{l\mu}(t,r) = \sum_{\rm poles}{c_{l\mu}(s_0) e^{s_0 t} f_l(s_0,r)} + \textrm {contour terms},
	\label{pole terms}
\end{equation}
where 
\begin{equation} \label{c}
	c_{l\mu}(s_0) \equiv \frac{1}{ dW_l/ds|_{s=s_0}}
	\int_0^{+\infty} {f_l(s_0,r') \mathcal{I}_{\, l\mu}(s_0,r') dx'}.
\end{equation}

Poles with $\Re(s_0)<0$ will correspond to exponentially damped 
oscillating-in-time quasinormal modes~\cite{reviewQNM}. (Despite 
the asymptotic behavior exhibited by $f^+_l (s_0,r)$ in 
Eq.~(\ref{bc1}), for $\Re(s_0)<0$, $\chi_{l\mu}(t,r)$ will be well
behaved at infinity due to the compact support initial condition
assumption~\cite{kay}.) Here, we focus on poles 
with $\Re(s_0)>0$ which will drive
$\chi_{l\mu}(t,r)$	to grow exponentially in time. The radial 
part of $\chi_{l\mu}(t,r)$ is determined by $f_l(s_0,r)$ 
as given by Eq.~(\ref{pole terms}) and satisfies Eq.~(\ref{feq}) 
with boundary conditions~(\ref{bc1}) and~(\ref{bc2}). Such a solution 
is a normalized eigenvector of the Hermitian operator $-d^2/dx^2 + 
V^{(l)}_{\textrm{eff}}(x)$, from which we conclude that 
unstable quasinormal modes have $\Im (s_0) = 0$ 
(see Fig.~\ref{fig:contour}). Let us denote by $\bar s_0$ the 
pole with largest positive $\Re(s_0)$ among them. Then, as far as 
our first order perturbation is valid, the ``late-time" behavior 
of $\chi_{l\mu}(t,r)$ can be cast in the form
\begin{equation} \label{unstableqnm}
	\chi_{l\mu}(t,r) \sim c_{l\mu}(\bar s_0) e^{\bar s_0 t} f_l(\bar s_0,r) 
	[1+ {\cal O} (e^{-\epsilon t})],
\end{equation}
where $\epsilon$ is some positive constant. Inserting this
expression in Eq.~(\ref{dphi}) and noting that the dominant
contribution comes from $l=0$, 
we  obtain 
\begin{align} \label{dphi2}
	\Phi_{(1)}^2 & \sim \left[ \int_0^{+\infty}{\!\!\!
	f_0({\bar s_0},r') \mathcal{I}_{\, 00}(\bar s_0, r') dx'}\right]^2
	\frac{e^{2\bar s_0 t}}{16 \pi \bar s_0^2} 
	\left[\frac{f_0(\bar s_0 ,r)}{r}\right]^2 
	 \nonumber \\
	&\times [1+{\cal O} (e^{-\epsilon t})],
\end{align}
where we have used $dW_l/ds|_{s=\bar s_0}= - 2 \bar{s}_0$
and that $f^\pm_l (s,r)$ can be rescaled arbitrarily  
to demand
\begin{equation}
\int_0^{+\infty}
	f_l^2 ( \bar s_0,r) dx =1.
\label{fnormalization}
\end{equation}
The expression for the 
Wronskian derivative can be obtained by adapting a derivation 
in~\cite{ching95} to real positive poles: 
first, we use Eqs.~(\ref{feq}) and~(\ref{fnormalization}) to write
\begin{align} 
1 & = \lim_{s\to \bar s_0}\int_0^{X \to \infty}  f_l^- (\bar s_0,r) f_l^+ (s,r) dx  
\nonumber \\
  & = \lim_{s\to \bar s_0}
\frac{
[f_l^+ (s,r)\partial_x f_l^- (\bar s_0,r)  
- 
f_l^- (\bar s_0,r)\partial_x f_l^+ (s,r)]_{r=0}
}{s^2-\bar s_0^2}, 
\label{Ws}
\end{align}
where the superior integration limit term vanishes as 
can be seen by using $\partial_x f_l^+ (s,r) \approx -s  f_l^+ (s,r)$
for $X$ large enough [see Eq.~(\ref{bc1})] and the fact that 
$f_l (\bar s_0,r)= f_l^\pm (\bar s_0,r) \stackrel{X\to +\infty}{\longrightarrow} 0$.
Then, by making the change $\bar s_0 \to s$ in the numerator of Eq.~(\ref{Ws})
to identify it with the Wronskian and using the L'Hospital rule, we 
obtain $dW_l/ds|_{s=\bar s_0}= - 2 \bar{s}_0$.

In order to compare the quantum and classical observables
$\langle 0_\textrm{in} |\hat \Phi^2 | 0_\textrm{in} \rangle$ and $\Phi_{(1)}^2$ given
by Eqs.~(\ref{evphi2}) and~(\ref{dphi2}), respectively, we
identify $\psi_{\bar{\Omega} 0}(r)$  and $f_0(\bar s_0 ,r)$
for $\bar \Omega=\bar s_0$, since 
they  satisfy the same differential 
equation [see Eqs.~(\ref{psieq2}) and~(\ref{feq})] 
with identical boundary conditions and compatible
normalizations.

Next, note that $\langle 0_\textrm{in} |\hat \Phi^2 | 0_\textrm{in} \rangle$ and  
$\Phi_{(1)}^2$ only differ by a multiplicative 
factor which includes $\mathcal{I}_{\, 00}(\bar s_0, r)$.
%\begin{equation}
%\frac{ \Phi_{(1)}^2} {\langle 0_\textrm{in} |\hat \Phi^2 | 0_\textrm{in} \rangle}
%= \frac{1}{2 \hbar \kappa \bar s_0}  
%\left[ 
%\int_0^{+\infty}{	f_0({\bar s_0},r') \mathcal{I}_{\, 00}(\bar s_0, r') dx'}\right]^2.
%\label{comparison}
%\end{equation}
This is natural since in the classical context the evolution of the scalar 
field depends on the choice of the initial conditions, while in  the quantum 
case the instability is triggered by vacuum fluctuations encoded 
on the choice of the quantum state. Now, let us suppose a compact star with 
radius $r=R_s$ and choose a typical initial condition as, e.g., 
$
\chi_{00}(t,r)|_{t=0} = A \Theta(R_s-r) 
$,
$
\partial_t \chi_{00}(t,r)|_{t=0} = 0
$.
In this case,
\begin{equation}
\frac{\Phi_{(1)}^2} {\langle 0_\textrm{in} |\hat \Phi^2 | 0_\textrm{in} \rangle}
= \frac{A^2 \bar s_0 }{2 \hbar \kappa}  
\left[ 
\int_0^{x(R_s)}{	f_0({\bar s_0},r')  dx'}\right]^2.
\label{comparisonpart}
\end{equation}
Finally, by using (i)~Eq.~(\ref{fnormalization}), (ii)~the fact that $f_0({\bar s_0},r)$
decreases fast for $r \gg R_s$, implying $f_0(\bar{s}_0,r) \sim 1/R_s^{1/2}$, and 
(iii)~$R_s^{-2} \sim |V^{(l)}_{\rm eff}| \sim \bar s_0^2$, 
we cast Eq.~(\ref{comparisonpart}) as
\begin{equation}
\Phi_{(1)}^2 /{\langle 0_\textrm{in} |\hat \Phi^2 | 0_\textrm{in} \rangle}
\sim {A^2}/{2 \hbar}.
\label{comparisonpart2}
\end{equation}
The consequence of condition (ii)~used above, namely $f_0(\bar{s}_0,r) \sim 1/R_s^{1/2}$,
comes by noting that $f_0(\bar s_0,r)$ gives a negligible contribution in 
Eq.~(\ref{fnormalization}) for $r\gtrsim R_s \sim x(R_s)$.
Condition (iii)~comes from Eq.~(\ref{feq}) by demanding that $|V^{(l)}_{\rm eff}|$
 be at least of order $R_s^{-2}$ to make the potential
 ``deep" enough to allow  bound solutions 
[see Eq.~(\ref{fnormalization})]. Conversely, by assuming the
existence of bound solutions the corresponding $\bar s_0^2$ is 
typically of the order of $|V^{(l)}_{\rm eff}|$.
\\

%%%%%%%%%%%%%%%%%%%%%%%%%%%%%%%%%%%%%%%%%%%%%%%%%%%%%%
\noindent
\textbf{Final discussions}:
We have shown how $\hbar$ can be made to appear in 
$\Phi_{(1)}$ by properly choosing the magnitude of the 
initial amplitude as $|A| \sim \hbar^{1/2}$. However, in 
this case a quantum mechanical treatment should be more suitable.
As long as fluctuations of the stress-energy-momentum tensor 
are ``reasonably" small~\cite{ford}, the spacetime will respond 
according to the semiclassical Einstein equations
$
G_{ab} = 8\pi \langle 0_\textrm{in} | \hat{T}_{ab} | 0_\textrm{in} \rangle
$.
The corresponding evolution is a highly nontrivial task. 
However, for unstable systems, it seems reasonable 
that when vacuum fluctuations become large enough
they should  somehow ``collapse" into classical perturbations 
(see, e.g., \cite{blencowe13}) in a process analogous to the formation of the 
cosmic-microwave-background anisotropies from primordial vacuum
fluctuations. Afterwards, the system should be properly evolved
through the classical equations of motion 
(see, e.g., \cite{novak,ruizetal}). 
%The transition from the quantum to the
%classical regimes involves subtle conceptual issues and is
%presently under investigation.
\\

%%%%%%%%%%%%%%%%%%%%%%%%%%%%%%%%%%%%%%%%%%%%%%%%%%%%%%
\noindent
\textbf{Acknowledgments}: 
We are thankful to Bob Wald for comments in the 19$^{\rm th}$ 
International Conference on General Relativity and Gravitation 
and in the 12$^{\rm th}$ Midwest Relativity Meeting which partially
motivated this paper. We also thank A. Landulfo and P. Pani for carefully
reading the manuscript. R. M. was supported by the S\~ao Paulo Research 
Foundation (FAPESP) under the Grant No. 2011/06429-3. G. M. and D. V. 
acknowledge partial support from FAPESP under Grant No. 2007/55449-1.
G. M. also acknowledges Conselho Nacional de Desenvolvimento 
Cient\'\i fico e Tecnol\'ogico (CNPq) for partial support.

%%%%%%%%%%%%%%%%%%%%%%%%%%%%%%%%%%%%%%%%%%%%%%%%%%%%%%
%\noindent
%\textbf{Appendix}:
\appendix*
\section{}
Here, we show that Eq.~(\ref{psieq2}) with $\xi = 0$ 
has no solutions satisfying conditions~(\ref{boundarycond}). 
It suffices to analyze the $l=0$ case because 
by vanishing the $l(l+1)/r^2$ positive term in the effective 
potential~(\ref{pot}), we improve our chances of finding 
bound solutions by ``deepening" $V^{(l)}_{\rm eff}$. 
Thus, we seek normalizable solutions of
\begin{equation}
	- {d^2\psi_{\Omega 0}}/{dx^2} + \left. V^{(0)}_{\textrm{eff}}\right|_{\xi =0} \psi_{\Omega0}
	 = -\Omega^2 \psi_{\Omega0},
\label{psiAppend}	 
\end{equation}
where
%\begin{equation} \label{potAppend}
$
\left. V^{(0)}_{\textrm{eff}}\right|_{\xi =0} = 
(e^{2(\Xi-\Lambda)}/{r}) 
\left( {d\Xi}/{dr} - {d\Lambda}/{dr} \right)
$
%\end{equation}
with
\begin{equation}
\psi_{\Omega 0}|_{x=0} =0
\label{boundary1}
\end{equation}
demanded by field regularity [see Eq.~(\ref{boundarycond})]. Furthermore,
the fact that $V^{(0)}_{\rm eff}|_{\xi=0}$ is nonsingular at the
origin demands $d\psi_{\Omega 0}/dx|_{x=0} = \textrm{const} \neq 0$.
For convenience, we choose  
\begin{equation}
d\psi_{\Omega 0}/dx|_{x=0} = C \exp [\Xi - \Lambda]|_{x=0}
\label{boundary2}
\end{equation}
with $C$ being a nonvanishing constant which is fixed by 
Eq.~(\ref{normpsi2}). We recall that Eqs.~(\ref{boundary1})-(\ref{boundary2})
uniquely determine the solutions of Eq.~(\ref{psiAppend}).

On the other hand, we see from Eq.~(\ref{x}) that $f(x)\equiv C r(x)$
satisfies (i)~the same differential equation as Eq.~(\ref{psiAppend}) 
provided that $\Omega=0$:
$
- {d^2 f}/{dx^2} + \left. V^{(0)}_{\textrm{eff}}\right|_{\xi =0} f = 0,
$
and (ii)~conditions similar to Eqs.~(\ref{boundary1})
and~(\ref{boundary2}), i.e.,
$
f(x)|_{x=0} =0 
$,
and
$
df /dx|_{x=0} = C \exp [\Xi - \Lambda]|_{x=0}
$.
Now, because $f(x)=C r(x)$ is a monotonically increasing function of $x$ 
[see Eq.~(\ref{x})], we immediately conclude that   Eq.~(\ref{psiAppend}) 
with $\Omega=0$ does not possess solutions satisfying Eq.~(\ref{boundarycond}). 
Then, because $\Omega^2$ in  Eq.~(\ref{psiAppend}) 
just increases $d^2 \psi_{\Omega 0}/dx^2$, we also get that the same 
conclusion is valid when $\Omega$ is nonzero. This implies that there are 
no unstable modes for minimally coupled scalar fields in asymptotically 
flat spherically symmetric static spacetimes containing no event horizons or
singularities, which is compatible with all known literature. 
Although the derivation above assumed a massless field,
the same conclusion holds for massive ones, $m\neq0$, since in this case 
the effective potential is altered by the addition of a positive 
term, $m^2 e^{2\Xi}$, which ``shallows" $\left.V^{(l)}_{\rm eff}\right|_{\xi=0}$ 
even more.

\end{document}